\documentclass[conference]{IEEEtran}
\IEEEoverridecommandlockouts
\usepackage{cite}
\usepackage{amsmath,amssymb,amsfonts}
\usepackage{algorithmic}
\usepackage{graphicx}
\usepackage{subcaption}
\usepackage{textcomp}
\usepackage{xcolor}
\usepackage[ruled,vlined]{algorithm2e}
\SetKwInput{KwInputa}{note event}    % Set the Input
\SetKwInput{KwInputb}{note events} 
\SetKwInput{KwInputc}{chord event}
\SetKwInput{KwInputd}{chord events}
\SetKwInput{KwInpute}{bounded}
\SetKwInput{KwInputf}{unbounded}
\usepackage{booktabs}
\usepackage{fancyhdr}

\usepackage{url}
\def\BibTeX{{\rm B\kern-.05em{\sc i\kern-.025em b}\kern-.08em
    T\kern-.1667em\lower.7ex\hbox{E}\kern-.125emX}}
    
\begin{document}

\title{Hierarchical Recurrent Neural Networks for Conditional Melody Generation with Long-term Structure\\

\thanks{This project is funded by Singapore's Ministry of Education under Grant no. MOE2018-T2-2-161 and SRG ISTD 2017 129.}
}

\author{
\IEEEauthorblockN{Guo Zixun}
\IEEEauthorblockA{\textit{Information Systems,}\\
\textit{Technology, and Design}\\
\textit{Singapore University}\\
\textit{of Technology and Design} \\
Singapore \\
nicolas\_guo@sutd.edu.sg}
\and
\IEEEauthorblockN{Dimos Makris}
\IEEEauthorblockA{\textit{Information Systems,}\\
\textit{Technology, and Design}\\
\textit{Singapore University}\\
\textit{of Technology and Design} \\
Singapore \\
dimosthenis\_makris@sutd.edu.sg}
\and
\IEEEauthorblockN{Dorien Herremans}
\IEEEauthorblockA{\textit{Information Systems,}\\
\textit{Technology, and Design}\\
\textit{Singapore University}\\
\textit{of Technology and Design} \\
Singapore \\
dorien\_herremans@sutd.edu.sg}

}

\maketitle

\begin{abstract}
The rise of deep learning technologies has quickly advanced many fields, including generative music systems. There exists a number of systems that allow for the generation of musically sounding short snippets, yet, these generated snippets often lack an overarching, longer-term structure. In this work, we propose CM-HRNN: a conditional melody generation model based on a hierarchical recurrent neural network. This model allows us to generate melodies with long-term structures based on given chord accompaniments. We also propose a novel and concise event-based representation to encode musical lead sheets while retaining the melodies' relative position within the bar with respect to the musical meter. With this new data representation, the proposed architecture is able to simultaneously model the rhythmic, as well as the pitch structures effectively. Melodies generated by the proposed model were extensively evaluated in quantitative experiments as well as a user study to ensure the musical quality and long-term structure of the output. We also compared the system with the state-of-the-art AttentionRNN~\cite{waite_2016}. The comparison shows that melodies generated by CM-HRNN contain more repeated patterns (i.e., higher compression ratio) and a lower tonal tension (i.e., more tonally concise). Results from our listening test indicate that CM-HRNN outperforms AttentionRNN in terms of long-term structure and overall rating. %The source code of the proposed CM-HRNN is available online \footnote{\url{https://github.com/AMAAI-Lab}}.
% The results show that the generated melodies are musically sound and contain more long-term patterns than AttentionRNN\cite{waite_2016}. The latter might be a first step towards enabling the generated music to elucidate the sense of earworms in music. 
% Models with a deeper hierarchy seem to generate music with a xxx structure.


\end{abstract}

\section{Introduction}
%intro
At the dawn of computing, the idea of generating music was first conceived by Lady Ada Lovelace, when she wrote: `[The Engine's] operating mechanism might act upon other things besides numbers [\dots] Supposing, for instance, that the fundamental relations of pitched sounds in the signs of harmony and of musical composition were susceptible of such expressions and adaptations, the engine might compose elaborate and scientific pieces of music of any degree of complexity or extent.'\cite{herremans2013composing}. When computers were first created, it was not long until they were used to generate the first piece of music \cite{hiller1958musical}. The field of computer generated music has grown ever since \cite{herremans2017functional}, with great strides being made in recent years due to deep learning technologies~\cite{briot2020deep}. Computer generated music, however, is not yet a part of our daily music listening experience. One potential reason could be the lack of repeating patterns or themes, i.e. longer-term structures, which are essential for the presence of `earworms' in music~\cite{burgoyne2013hooked}.
%audio&symbolic domain generation
There are two different types of music generation systems: those that generate symbolic music and those that generate raw audio. Some attempts have been made to generate music directly as raw audio \cite{mehri2016samplernn,oord2016wavenet}. This remains a challenge, however, due to the disproportionate amount of audio samples in the overall musical structure. Hence, lots of of the existing music generation research focuses on the symbolic domain where music is represented as a series of musical events in sequence. Popular data encoding schemes for symbolic music include piano-roll representation~\cite{yang2017midinet,eck2002first}, tonnetz representation \cite{chuan2018modeling}, word embeddings \cite{huang2016chordripple,chuan2020context}, and event-based representation \cite{oore2020time,makris2021seq}. In this work, we will focus on the symbolic approach, more specifically, we will generate melodies from their respective chord accompaniments using a novel event-based representation. Compared to the existing event-based representations, we explicitly encode bar event information in our representation. This allows our model to understand musical complexities such as meter \cite{de2016meter}.

Composing melodies from a given set of chords is a task faced by many musicians in the real world, e.g. in pop music composition or jazz improvisation. We are thus motivated to design our model to receive a chord sequence as input conditions by the users and generate melodies based on these provided chords. This way, the users will have a certain level of control over the generated melodies through the manipulation of the input chord sequence.

% Our proposed system tackles this task by generating melodies that match a chord sequence inputted by the user. We thus propose a computer-assisted-composition tool, which allows the user a certain level of control over the generated melody through the manipulation of the input chord sequence.

Since music data is sequential, researchers have recently focused on developing recurrent neural networks (RNN) and their variants (e.g., long-short term memory (LSTM) / gated recurrent units (GRU)) for music generation due to their memory mechanism \cite{briot2020deep}. Still, current computer generated music often lacks a long term structure, an essential quality of polished, complete musical pieces with recurring themes \cite{herremans2015generating,huang2018music,herremans2017functional}. Even if it is speculated by researchers that the generated music may include long-term structure, very rarely a quantitative measure is employed to measure this~\cite{chuan2018modeling}. By injecting a hierarchical structure between the RNN layers of our proposed model, we allow the model to capture musical patterns on different time scales. In this work, we propose CM-HRNN, a conditional melody generation model based on a hierarchical recurrent neural network. In Section \ref{sec:data_repres} and \ref{sec:model}, we describe the proposed event representation and CM-HRNN architecture in detail. We then thoroughly analyze the music generated by CM-HRNN in terms of musical quality and the presence of repeated patterns in Section 
\ref{sec:experiment}. The latter is achieved by calculating the compression ratio of the generated melodies\cite{meredith2002algorithms}. Finally, the model output is evaluated in a listening study. In the next section, we first give an overview of the related work. 

\footnote{Preprint accepted for publication in the Proc. of the International Joint Conference on Neural Networks (IJCNN), Shenzhen, China, 18-22 July 2021(virtual).}

\section{Related work}
%audio domain: samplernn, wavenet%
%sliding window: melodyrnn, lookbackrnn, attentionrnn%
%hierachy: HRNN, music VAE, song from pi%
%attention mechanism: transformer%

We will start by giving a brief overview of existing music generation systems with a focus on generating long-term structures. Then we will dive into hierarchical models which have shown to be effective in capturing long-term dependencies in multiple fields such as natural language processing \cite{chung2016hierarchical}, audio generation \cite{mehri2016samplernn,koutnik2014clockwork}, and music generation \cite{wu2019hierarchical,roberts2018hierarchical}.

\subsection{Music generation systems with long-term structure}

The problem of generating music with long-term structure has received limited attention. The affective music generation system MorpheuS \cite{herremans2017morpheus} enforces repeated patterns and structure into music by considering music generation as a combinatorial optimization problem. MorpheuS uses repeated patterns from existing templates as hard constraints during generation. Looking towards machine learning methods, \cite{herremans2015generating} use a Markov model that learns the statistical properties of a musical corpus, and combines this with a variable neighborhood search optimization algorithm to generate music, while hard constraining a larger structure. Similarly, \cite{lattner2018imposing} train a convolutional restricted Boltzmann machine (C-RBM), but use simulated annealing as a sampling technique. This allows them to include structural constraints.

%before this, optimization combined methods. Below this, pure NN models
Looking at ``pure'' neural networks that do not leverage slower optimization techniques, \cite{chuan2018modeling} use a novel Tonnetz representation to train an LSTM that is better able to generate polyphonic music with repeated patterns than a similar network with a more traditional piano roll representation. 
Two recent RNN-based systems, LookbackRNN and AttentionRNN \cite{waite_2016}, aim to generate monophonic melodies with long term structure in an autoagressive way. Both models incorporate the ``repetitive label'' in the data representation, a way to represent repetition of notes in the neighbouring two bars.
%<THINK OF SOME SEQ-SEQ MODELS>.Seq2seq models like transformers(\cite{}) and music transformer(\cite{}) could generate music from external conditions.
%
LookbackRNN \cite{waite_2016} contains a two-layer LSTM network with residual connections between different time steps. This allows the model to feed note events from the previous 1 to 2 bars to the current time step. The residual connection is able to offset the impact of vanishing gradients and allows the model to generate more repetitive patterns. Similar to LookbackRNN, AttentionRNN \cite{waite_2016} also inputs previous event information into the current generation step, however, with a learnable attention mask applied to the previous $n$ events. The attention mask decides how much attention to put on each of the previous events. By explicitly feeding the neighbouring events into the current generation step, the two models are able to better learn when recent melody repeats. Each generation step, however, may not be aware of ``the bigger picture''. 

% To address this, we condition each of our generation sliding windows with the fine-to-coarse context from upper RNN layers in our proposed model. 

Another model is the MusicVAE \cite{roberts2018hierarchical}, which generates smooth transitions between two musical fragments with overall long-term structures. The input music data is first encoded by a bidirectional RNN encoder which generates a latent vector $z$. The latent vector $z$ will then be fed as a initial state to an RNN which generates a series of ``conductors'' which will then ``conduct'' the hierarchical RNN decoder to generate music sequences. Since each ``conductor'' conditions the generation for multiple steps, the generated music is shown to contain more long-term structures compared to other existing recurrent variational autoencoders. Even though the problem domain is different, we can still gain inspiration from the conditioning ``conductors'' and the hierarchical decoders which enhance the overall long-term structure in music.

Hierarchical RNN (HRNN) \cite{wu2019hierarchical} aims to generate monophonic melodies with long-term structure. It first generates bar profiles and beat profiles which indicate the rhythmic pattern within a bar and a beat using vanilla LSTMs. Beat profile generation is directly conditioned by the latent output from the upper RNN layer which generates the bar profiles. These two profiles then jointly condition the pitch generation. By combining the generated pitch and rhythmic pattern, the generated monophonic melodies are qualitatively evaluated through multiple listening tests and are shown to outperform the LookbackRNN \cite{waite_2016}. The core design of the architecture is to first generate the rhythmic patterns and then condition the pitch generation with both coarse and fine rhythmic patterns. 

In this paper, we propose an architecture similar to HRNN but with different design motivations. We believe that rhythmic and pitch features are equally important and hierarchical structures should exist in both domains. Hence, instead of conditioning the pitch generation with the generated rhythmic patterns, we apply the hierarchical RNN structure to both rhythmic and pitch latent spaces simultaneously. Compared to HRNN, our proposed CM-HRNN is able to condition the melody generation with the provided chords. Moreover, the musical quality of HRNN-generated music is only evaluated through a subjective listening test. In contrast, we evaluate our proposed CM-HRNN with extensive analytical measures as well as a user study.

\subsection{Hierarchical architectures for long term dependencies}

Drawing inspiration from other fields in which long-term dependencies of sequential data have been modelled by RNNs, we find that multiple hierarchical architectures have been proposed for this challenge\cite{schmidhuber1992learning,hihi1995hierarchical,koutnik2014clockwork,chung2016hierarchical, huzaifah2020mtcrnn}. These models try to control the weight update rate within the RNN cells. RNN weights matrices are separated into different chunks and are updated using different rates: high update rates capture the short-term dependencies whereas low update rates capture long-term dependencies. 

These architectures can be applied to the domain of audio generation. For instance, SampleRNN\cite{mehri2016samplernn} is able to synthesise realistic sounding audio with a hierarchical RNN structure. Instead of manipulating the RNN weights update rate to capture the temporal and long-term structure of sequences of audio samples, it uses multiple stacks of RNN with upper tier RNN layers operating on more grouped audio samples per step and lower tier RNN layers operating on fewer grouped samples per step. Inspired by SampleRNN, we group our data (i.e., music events) in different resolutions and process these groups of data separately to obtain the coarse-to-fine hierarchical features of the data.

%  Compared to other types of sequential data like text or raw audio, symbolic music data contains unique types of structures and hierarchies. For example, pop music usually contains sections like verse or chorus, each section contains multiple phrases and each phrase may contain multiple repetitions and modulations. We will review existing research which focuses specifically on modelling these long-term structures of music data.

% These models are proposed to capture long term structures of sequential data in general, yet, none of these architectures are used to model symbolic music data, which contains unique types of structures and hierarchies compared to other sequential data like text or raw audio. For example, pop music usually contains sections like verse or chorus, each section contains multiple phrases and each phrase may contain multiple repetitions and modulations. Hence, instead of manipulating the weight of the RNN, we decide to opt for a different method inspired by the architecture of SampleRNN\cite{mehri2016samplernn}: using multiple stacks of RNN with upper tier RNNs operating on coarsely grouped data and lower tier RNN operating the opposite. In this manner, without changing the weight updating rules, RNN cells from different tiers will capture dependencies in both fine and coarse resolution.

\section{Event-based representation with understanding of meter}\label{sec:data_repres}
% Inspired by \cite{colombo2016algorithmic}, we propose a novel data encoding scheme. 

We propose a novel data encoding scheme based on \cite{colombo2016algorithmic} which uses two combined one-hot encoded sub-vectors to represent the pitch and duration of music events. Additionally, we extend their representation with another three sub-vectors. As a result, each music event vector consists of the following sub-vectors: 

\begin{itemize}
    \item \textbf{Pitch:} The MIDI standard defines 128 pitches. We add two additional elements to this sub-vector: one to indicate a rest; and one to indicate if a note is sustained or not (i.e., tied note). This results in a 130-dimensional sub-vector.
    \item \textbf{Duration:} A 16-dimensional vector which represents duration. Possible values range from a 16th note to a whole note with an increment of 16th note duration (i.e. smallest quantization unit).
    % reflects the note duration, expressed in terms of sixteenth (16th) notes (smallest unit). 
    \item \textbf{Current and Next Chord:} There are 12 possible chord roots (from C to B). For each root we set 4 possible chord types (i.e. major, minor, diminished, and dominant 7th) plus the rest symbol. Since the chord information is used to guide the melody generation, both the current chord as well as the next chord are provided as input to the network at each time step. This results in two 49-dimensional vectors. 
    \item \textbf{Bar:} A two-dimensional vector which indicates if it is the start of a bar or not. 
\end{itemize}

\begin{figure}[htbp]
 \centering
\includegraphics[scale=0.27]{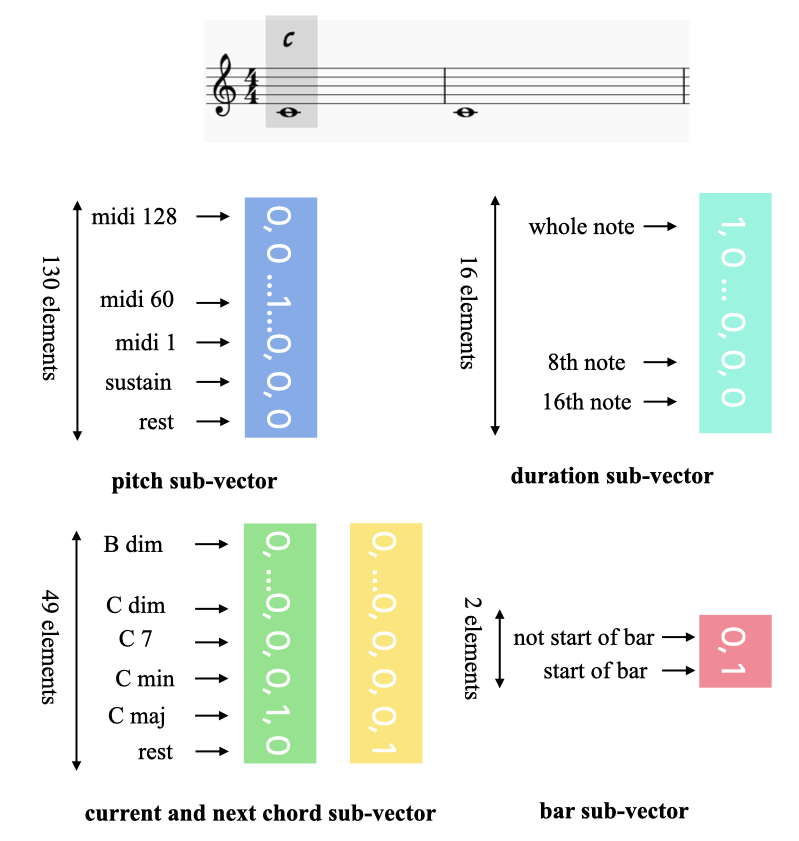}
 \caption{Example of the sub-vectors for the starting C note.}
 \label{fig:data_repres1}
 \end{figure}

\begin{figure}[htbp]
 \centering
\includegraphics[scale=0.20]{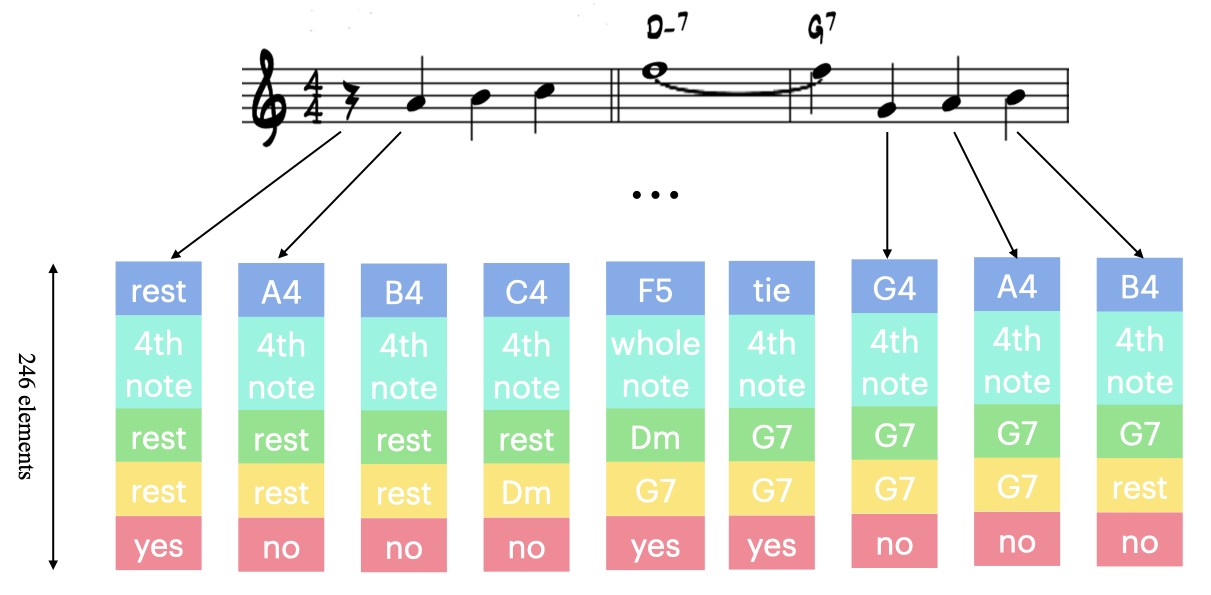}
  \scriptsize
 Each color-coded block denotes a one-hot vector. For illustration purposes, only the event symbol is listed in the color-coded block. Readers can refer to the sub-vector with the same colour in Fig. \ref{fig:data_repres1} for the corresponding one-hot vector representation.
 \caption{Proposed event-based representation for the first three bars of ``autumn leaves''.}
 \label{fig:data_repres}
 \end{figure}

An example of these sub-vectors is shown in Fig. \ref{fig:data_repres1}. All sub-vectors are concatenated together. This inclusion of multiple types of data, or viewpoints, such as bar information, is inspired by \cite{conklin1995multiple}. Finally, an example of the proposed lead sheet encoding is shown in Fig. \ref{fig:data_repres}.

%As a result, each music event vector consists of five sub-vectors: one to represent the pitch, duration, current (underlying) chord, next chord, and bar. All sub-vectors are one-hot encoded and are concatenated together. An example of these sub-vectors can be found in Figure~\ref{fig:data_repres1}. The inclusion of multiple types of data, or viewpoints, such as bar information is inspired by \cite{conklin1995multiple}.

%The pitch sub-vector consists of 130 elements, one element for each of the 128 midi pitches, one element to indicate if it is a rest, and one element to indicate if the note is sustained or not (i.e., tied note). The duration sub-vector consists of 16 elements, the first element represents a 16th note duration and the last a whole note duration. The elements in between are incremented by 16th-note durations.
%The third sub-vector, i.e. chord vector, consists of 49 elements. The first element indicates a rest symbol. For the remaining 48 elements, each element represents a combination of the tonic center of the chord with one of the four possible chord types. There are 12 tonic centers (starting from C to B). Each tonic center has four possible chord types: major, minor, diminished, and dominant 7th. Since the chord information is used to guide the melody generation, both the current chord as well as the next chord will be provided. Finally, the bar sub-vector consists of two elements. Each element represents whether the current event is at the start of a bar. In Fig.~\ref{fig:data_repres}, three bars of ``Autumn Leaves'' are used as an example to illustrate the proposed encoding. 

Since bar lines are included in the proposed notation, we can calculate the accumulated time information for each event within the bar as shown in Table \ref{tab:acc_tm} which will then be fed into our proposed model in the bottom generation tier. Explicitly feeding the accumulated time information into to model is shown to improve the duration and bar event prediction (see Section \ref{sec:experiment}).
% The accumulated time information will then be fed to our proposed model to make it aware of the relative positioning information. The additional input will later be proved to be effective in terms of duration and bar event prediction in Section \ref{sec:experiment}.

\begin{table}[h!]
\caption{Accumulated time information for the first three bars of ``autumn leaves''.}
\centering
\begin{tabular}{ cccccccccc } 
\toprule
Event number&1 & 2 & 3 & 4 &5 &6 & 7 & 8 & 9 \\  \midrule
Start of bar &yes & no & no & no &yes &yes & no & no & no\\
Note duration &1 & 1 & 1 & 1 &4 &1 & 1 & 1 & 1\\ 
$acc\_t$ &1 & 2 & 3 & 4 &4 &1 & 2 & 3 & 4\\ 
% event number&1 & 2 & 3 & 4 &5 &6 & 7 & 8 & 9 \\  \midrule
% dur &4th note & 4th note & 4th note & 4th note &4 &4th note & 4th note & 4th note & 4th note\\ 
% acc\_tm &1 & 2 & 3 & 4 &4 &1 & 2 & 3 & 4\\ 
% if bar &yes & no & no & no &yes &yes & no & no & no\\ 
\bottomrule

\end{tabular}
\\
\smaller
\smallskip
Duration is expressed in units of one quarter note here.  %``dur'' stands for duration in 16th notes and 
``acc\_t'' is accumulated time.
\label{tab:acc_tm}
\end{table}

There are several advantages to this representation. Firstly, pitch repetition and duration repetition are mutually inclusive (i.e. they can co-exist or occur separately)\cite{medeot2018structurenet}. For example, a melodic motif may be repeated several times in a jazz piece with slight rhythmic variations. In other cases, repetitive patterns may exist in the duration vector space but not in the pitch vector space. Hence, using multi-hot encoded data enables RNNs to learn different high-level features and patterns separately. Secondly, this representation can encode data efficiently. Compared to the piano-roll representation, our proposed encoding can easily represent two consecutive notes with the same pitch whereas one may not be able to differentiate between a long note or two repeated notes in the piano-roll representation~\cite{eck2002first} unless additional onset information is added to capture the repetitive onset which could potentially increase the computational cost. Moreover, given that the smallest duration value used(i.e. the quantization step) is a 16th note in this research, a piano-roll representation would need 32 event vectors to represent two tied whole notes, whereas our proposed representation only needs 2 event vectors: [whole note with a duration of 4 beats, tied note with a duration of 4 beats]. Thirdly, it allows us to incorporate an indication of bar events, which is crucial to musical composition and will allow the system to learn the relative positioning within bars. Last but not least, the proposed data representation could easily be extended to include other features (e.g., velocity). This is a major difference between our work and \cite{colombo2016algorithmic}.

\begin{figure*}[h!]
     \centering
        \includegraphics[scale=0.25]{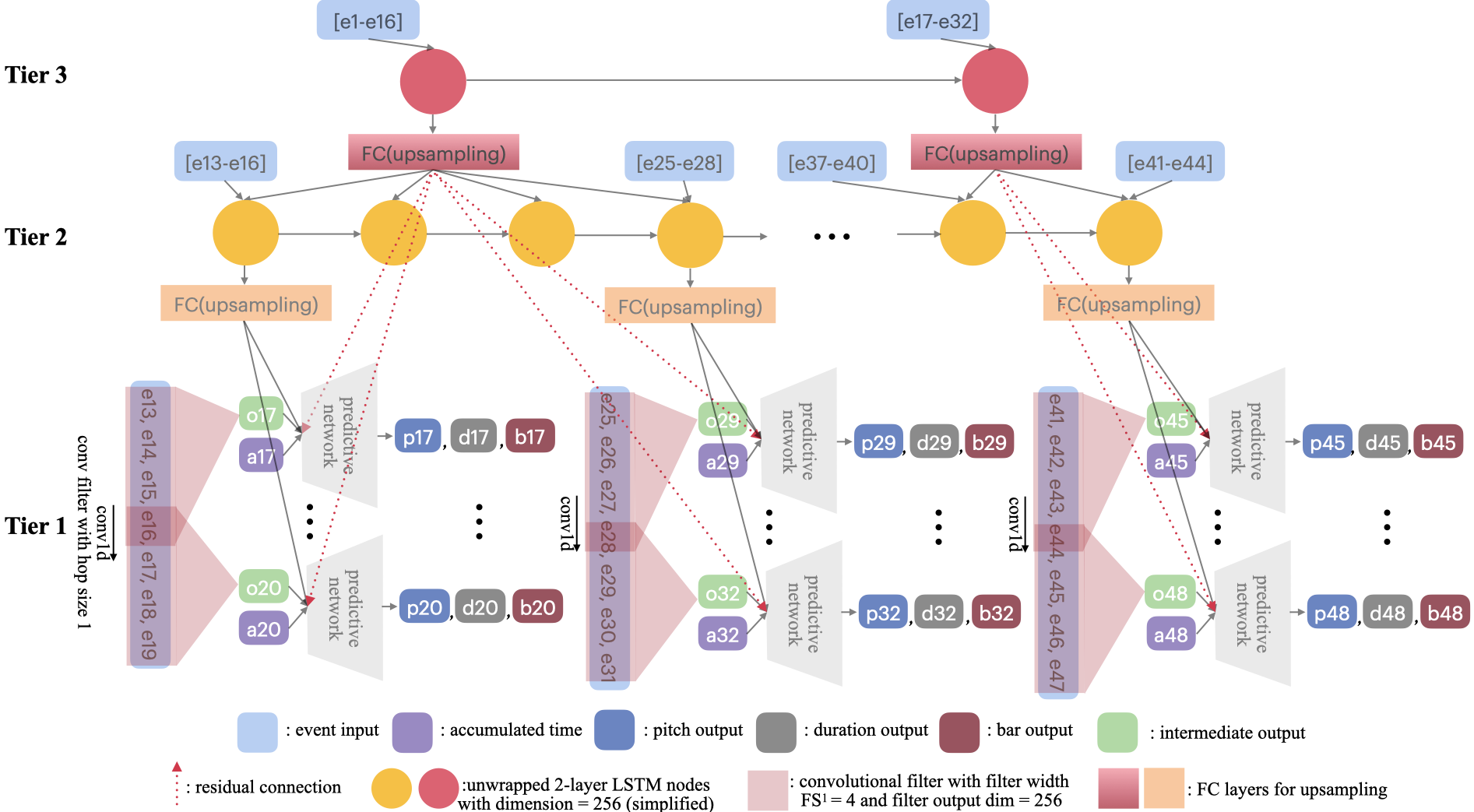}\\
  \caption{Our proposed CM-HRNN architecture.}
    \label{fig:architecture3tu}
\end{figure*}

\section{Proposed model: CM-HRNN}\label{sec:model}

\subsection{Model Architecture}

\begin{figure}[h!]
     \centering
        \includegraphics[scale=0.35]{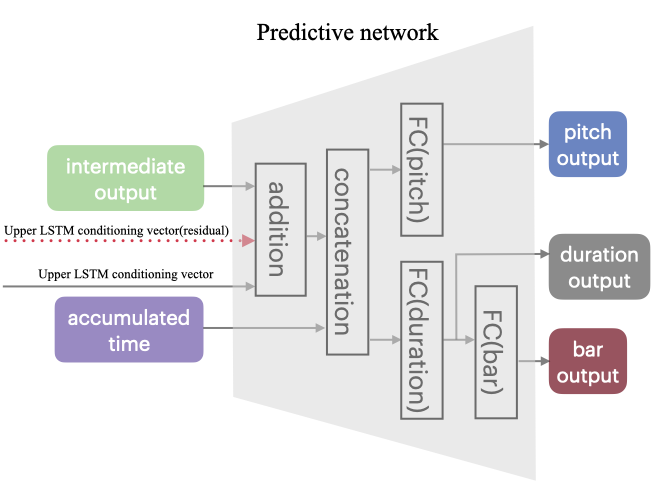}\\
  \caption{Details of the predictive network within CM-HRNN. FC(pitch) represents 2 fully connected layers (130 nodes each). FC(duration) represents 2 fully connected layers (16 nodes each). Finally, FC(bar) is a 1-layer fully connected network. All FCs have ReLu activation. }
    \label{fig:pred_network}
\end{figure}

Our proposed CM-HRNN architecture consists of multiple hierarchical tiers (see Fig. \ref{fig:architecture3tu}). All tiers except for the bottom tier use LSTM cells to process grouped event data. Upper tiers process more grouped event data per step to capture the long-term dependencies. The latent output from the upper tiers, which represents a coarser latent representation of sequential data, directly conditions the lower tiers' generation after proper upsampling. The bottom tier, after receiving fine-to-coarse conditioning vectors from the upper tiers, proceeds to predict future music events. Inspired by \cite{mehri2016samplernn}, the CM-HRNN's bottom tier uses a conv1d operation to process overlapping sliding windows of $n_{tier_1}$ events ($1$ indicates the bottom tier) and generate the next event. To convert the latent output from these sliding windows into music events, we include a predictive network in the bottom tier (see Fig. \ref{fig:pred_network}). In this paper, we focus on two variants of the proposed architecture: a 2-tier CM-HRNN and a 3-tier CM-HRNN. In theory, we could incorporate as many tiers as possible, but to avoid overly increasing the model's complexity, we will restrict the number of tiers to a maximum of three. The architecture of the proposed 3-tier CM-HRNN is shown in Fig. \ref{fig:architecture3tu}. By removing the top tier of the 3-tier CM-HRNN, we obtain the architecture of the 2-tier CM-HRNN.

In the 3-tier CM-HRNN, residual connections exist between the top tier and the bottom tier to alleviate the effect of the vanishing gradient. This also allows the bottom generation tier to receive the latent conditioning output from all of the upper tiers directly. Additionally, we also feed the accumulated time information to the predictive network in the bottom generation tier. The accumulated time vector has the same representation as the duration vector (see Section \ref{sec:data_repres}, Table \ref{tab:acc_tm}). Given that the position of each bar line is indicated in our proposed data representation, the model is able to quickly learn the relationship between the accumulated time information, the duration vector, and the bar event vector, which makes the model more aware of relative positioning information.

% \begin{figure}[htbp]
%  \centering
% \includegraphics[scale=0.25]{figures/accumulated_time_info.png}
%  \caption{Accumulated time information}
%  \label{fig:acc_t_info}
%  \end{figure}
 
In the upper tiers of the model, events are grouped into non-overlapping event frames. Each such frame contains $FS^k$ events. Whereas in the bottom tier, events are grouped into overlapping event frames with a stride of 1 (i.e., sliding windows). Here, $k$ indicates the tier number and $FS^k$ represents the frame size of tier $k$. These event frames will be processed by their respective tier in order. The relationship between each $FS^k$ and tier is as follows:

% \begin{align}
% n_{tier_1} &= n_{tier_2}\\
% n_{tier_{k+1}} mod\; n_{tier_k} &= 0\\
% n_{tier_{k+1}} &> n_{tier_k}
% \end{align}

% $n_{tier_k}$

\begin{align}
FS^{k=1} &= FS^{k=2}\\
FS^{k+1} mod\; FS^k &= 0\\
FS^{k+1} &>  FS^k &&\text{if } k > 1
\end{align}

We group events into event frames to form the input for the different tiers. Here, $e_t$ indicates the $t$-th event. All $e_t$s within one square bracket are grouped into one event frame.

\begin{align}
    frames_k&= 
        \begin{cases}
        [e_1...e_{FS^k}],[e_{FS^k+1}...e_{2\cdot FS^k}]...,& \text{if } k \neq 1  \\
        [e_1...e_{FS^k}],[e_2...e_{FS^k+1}]...,& \text{if } k = 1
        \end{cases}
\end{align}

%Dorien: next section should be checked thoroughly, or some equations removed if you can refer to original paper. 
We define the input for each tier's processing unit (2-layer LSTM or conv1d) $i_k$ as follows:
\begin{align}
    i_k&= 
        \begin{cases}
        frames_k,   & \text{if } k = top\; tier \;\text{or}\; 1  \\        
        W_fframes_k+W_{o_i} \boldsymbol{o_{k+1}}, & \text{otherwise}
        \end{cases}
\end{align}
Different tiers receive input at different rates, hence, the output from the upper tier LSTMs needs to be upsampled before it is used as a condition in the lower tier generation. In the formula below, $h$ indicates the hidden states of the LSTM; $o_{t_k}$ indicates the intermediate output for tier $k$ at the $t$-th time step; $\mathbf{o_{t_k}}$ represents the upsampled LSTM output; $acc_t$ represents the accumulated time information at time step $t$; $\oplus$ is used for vector concatenation; and all $W$s are trainable weights. Bias terms and repeated layers are omitted in the equation for simplicity. Residual connections between tiers are implemented as per \eqref{eq:res}.

\begin{align}
    o_{t_k},h_{t_k}&=\boldsymbol{LSTM}(i_{t_k},h_{{t-1}_k}),&&\text{if k} \neq1\\
    \boldsymbol{o_{t_k}} &= W_{o_{\text{upsample}}}o_{t_k}, &&\text{if } k \neq top\; tier\;\text{or}\;1\\
    \label{eq:res}o_{t_1} &=(W_ii_{t_1}+  \sum_{k=2}^{k_{top}} o_{t_k} )\oplus acc_{t} 
\end{align}

% \subsection{Loss definition}

To predict the next event at the $t+1$-th time step, we input the intermediate output $o_{t_1}$ to the predictive network as per Fig.~\ref{fig:pred_network}. Variables $p_t$, $d_t$, and $b_t$ represent the pitch, duration, and bar sub-vectors of an event respectively. Equations \eqref{eq:pred_start} to \eqref{eq:pred_end} represent the predictive network, whereby $FC()$ can consist of multiple dense layers (as per Fig.~\ref{fig:pred_network}) with ReLu activations. For simplicity, we did not include the activation information in the equations. We then apply three softmax functions to obtain the probability distribution for each of these three sub-vectors. 
\begin{align}
    \label{eq:pred_start}p_{t} &= FC_{p}(o_{t_1})\\
    d_{t} &= FC_{d}(o_{t_1})\\
    \label{eq:pred_end}b_{t} &= FC_{b}(d_t)\\     
    \hat{pitch_t} &\thicksim Softmax(p_t)\\
    \hat{duration_t} &\thicksim Softmax(d_t)\\
    \hat{bar_t} &\thicksim Softmax(b_t)
\end{align}

We calculate the cross-entropy loss for each of the three vectors (pitch, duration, and bar) and sum them after given them different weights $\alpha_1$, $\alpha_2$, and $\alpha_3$, respectively, to form the final training objective function. By assigning different weights to each of the vectors, we are able to emphasise each loss term based on its importance.
\begin{align}
    Loss = &{\alpha_1}\cdot CE(\hat{pitch_t},pitch_t) \\
           +&{\alpha_2}\cdot CE(\hat{duration_t},duration_t)\nonumber\\
           +&{\alpha_3}\cdot CE(\hat{bar_t},bar_t)\nonumber
\end{align}

% We further augmented this model by adding residual connections in the xxx layer. This allows the network to better... 

\subsection{Implementation details}

We have implemented several models which will be evaluated in Section \ref{sec:experiment}. All models utilise a two-layer LSTM with 256 nodes as the memory unit. The sliding window in the bottom tier is implemented as a conv1d operation with a stride of 1. The filter for the convolution has a filter size of $FS^1$ (varies per experiment) and an output dimension of 256. In the predictive network, we use 2 fully connected (FC) layers (130 nodes each) for pitch prediction; 2 FC layers (16 nodes each) for duration prediction; and 1 FC layer (2 nodes each), built upon the duration prediction network, for bar prediction. All FC layers have ReLu activation. Through trial-and-error, we found that setting $\alpha_1=0.4$, $\alpha_2=0.3$, and $\alpha_3=0.3$ yields the best output quality. 
To generate music for our experiments, we provide the first 16 pitch events and all chord events of all songs in the test set as the model input. The sampling temperatures for the pitch and duration and bar output distribution are set to be 0.7, 0.2, and 0.1 respectively. The duration sampling temperature is set low whereas the pitch sampling temperature is set high because we want to keep the timing correct while giving more freedom to pitch generation.

\section{Experiments}\label{sec:experiment}

\subsection{Experimental setup}
We set up several experiments to determine the optimal network architecture; evaluate the model's ability to generate high-quality music with structure, and to compare it with a state-of-the-art system, AttentionRNN. In a first experiment, we tested whether the accumulated time information ($acc_t$) could help the model when generating bar events and thus better understanding the meter. Hence, we implemented 2 sets of models: with and without the accumulated time information ($acc_t$) fed as input. Next, we tested the influence of the residual connection and the frame size $FS^2$ in the 3-tier CM-HRNN, on the model's ability to include repeated patterns in the generated music, as well as keep track of the musical meter. Finally, the best 2- and 3-tier CM-HRNN are compared to AttentionRNN, both analytically, as well as through a listening test. While it is not always easy to compare different music generation systems, given that they often have slightly different functional tasks or input representations, a comparison with AttentionRNN was facilitated through their use of multi-hot encoded input data which is similar to our data representation. Below, we first describe our dataset and pre-processing method, followed by a description of the evaluation metrics and the listening test setup.

\subsection{Dataset and pre-processing}
%theorytab, xml
%analyse dataset, how many songs per genre; data format(chord, melody)
%choose time signature 4/4, in C, only monophonic
%merge intro, verse, pre-chorus, chorus,outro in sequence(make sure long sequences)
%in absolute time, changed into delta time 
%how we process dataset:algorithm
All training data (XML format) was parsed from Theorytab~\cite{TheoryTab}, an open-source website that provides lead sheets. Each XML file contains a label with the song genre, song sections, note events, and chord events with absolute timing. During the data encoding stage, we changed the absolute timing into delta timing due to the nature of the LSTM cells. For simplicity, we included only monophonic melodies with a time signature of 4/4, along with their chords. All songs were transposed to C. We merged all available sections (i.e., intro, verse, pre-chorus, chorus, and outro) of the same song to make the training sequences long enough. We matched each note event with its corresponding chord and labeled the note event sequences with the bar label. In the end, we obtained 5,507 musical fragments which were then randomly split into training, validation, and test set with the ratio of 0.8, 0.1, and 0.1 respectively.

\subsection{Analytical evaluation measures}

\subsubsection{Compression ratio to measure long-term structure}
We use COSIATEC\cite{meredith2002algorithms} as a proxy to evaluate the long term structures contained in the generated music as in \cite{chuan2018modeling}. COSIATEC utilizes a geometric approach much like zip-file compression, to detect repeated patterns in symbolic music data. The resulting compression ratio indicates how much smaller a musical file can be made by representing it with pattern vectors and their corresponding translation vectors. %For example, if a pattern vector has $3$ datapoints (i.e., notes) and it occurs $3$ times in a piece, the total repeated patterns could be represented by $1$ position vector of length $3$ and $2$ translation vectors (two coordinates: delta time and pitch shift). As a result, $9$ datapoints are in effect represented by $6$ data points, achieving a compression ratio of $2$ assuming the piece consists of no other notes. 
Hence, this compression ratio reflects the number of repeated patterns that the generated music contains. In the experiments, we only calculate the compression ratio of the generated \emph{melodies}.

\subsubsection{Tension measures}
We use the model for tonal tension\cite{herremans2016tension} based on the spiral array \cite{chew2002spiral}, to measure the amount of tension in the generated music. This model offers three measures for each music time window (i.e., cloud): cloud diameter, tensile strain, and cloud momentum. Cloud diameter indicates the tonal dissonance within a cloud, tensile strain measures the tonal distance to the key of the song, and cloud momentum calculates the tonality movement between different clouds. Even though a limited amount of dissonant notes could sound more musically interesting, these could also adversely affect the music quality. By calculating these tension measures, we have a direct impression of whether the generated music is tonally consistent.

\subsection{Listening test setup}

An online listening test was conducted to evaluate the proposed model subjectively. Each participant was asked to listen to 12 musical fragments ranging from 13 to 44 seconds generated by 3 models: 2-tier CM-HRNN, 3-tier CM-HRNN, and AttentionRNN (with an attention size of 32). These snippets of music are placed in random order and the lead sheets were shown during playback. All participants were asked to rate the following questions on a scale from 1 (very poor) to 5 (very good): 
% 1. the overall perception of musical quality; 2. the coherence of the music; 3. the consonance between chord and melody; 4. the naturalness of the generated melody in terms of pitch; 5. the naturalness of the generated melody in terms of duration.

\begin{enumerate}
    \item overall perception of musical quality;
    \item coherence of the music; 
    \item consonance between chord and melody;
    \item naturalness of the generated melody in terms of pitch;
    \item naturalness of the generated melody in terms of duration.
\end{enumerate}

\section{Results}

\subsection{Ability to learn correct bar timing and effectiveness of the accumulated time information}

\begin{figure}[t]
	\centering
	\begin{subfigure}{.5\columnwidth}
		\centering
		\includegraphics[width=\linewidth]{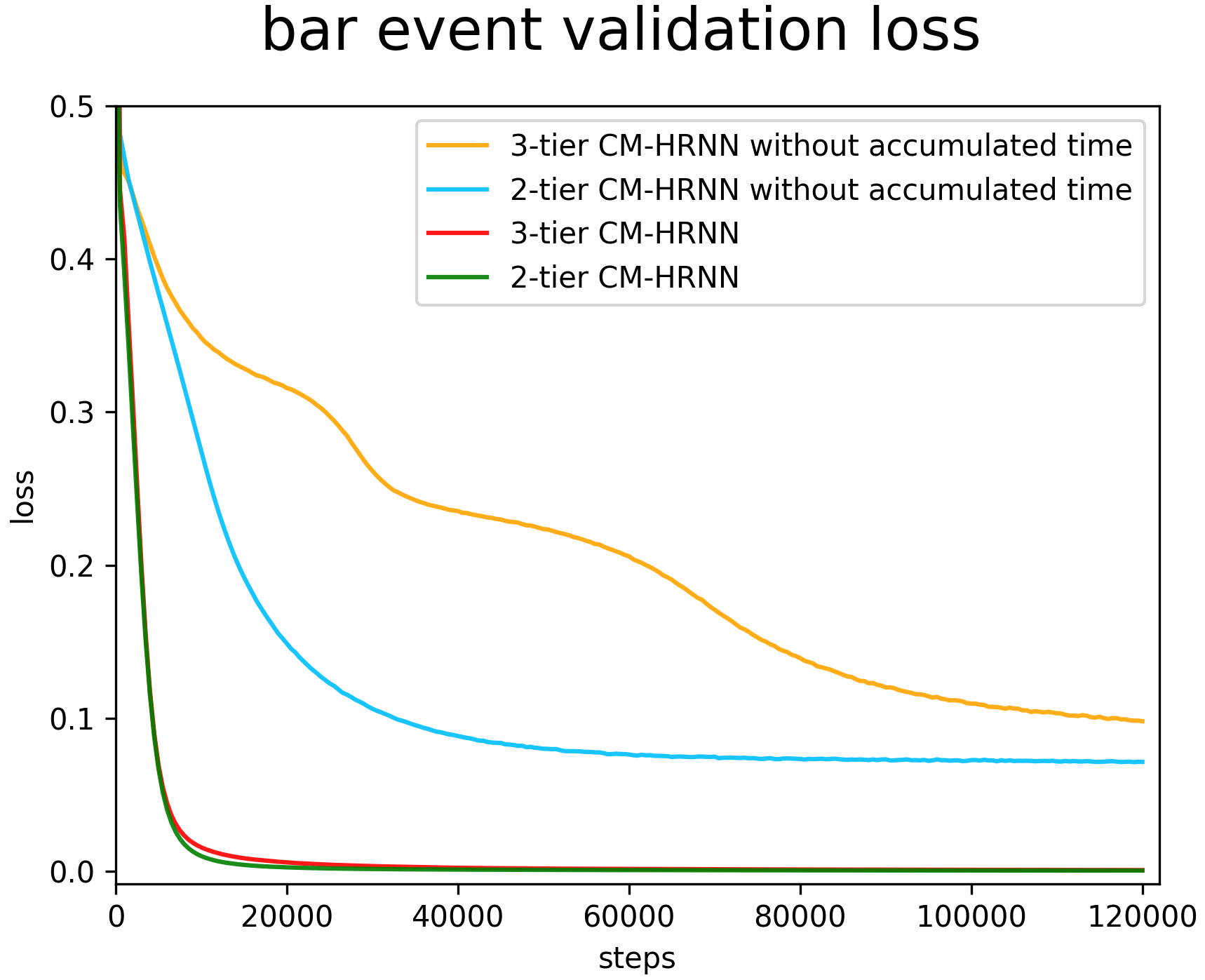}
		\caption{Bar event validation loss.}\label{fig:1a}		
	\end{subfigure}%
  \hfill
	\begin{subfigure}{.5\columnwidth}
		\centering
		\includegraphics[width=\linewidth]{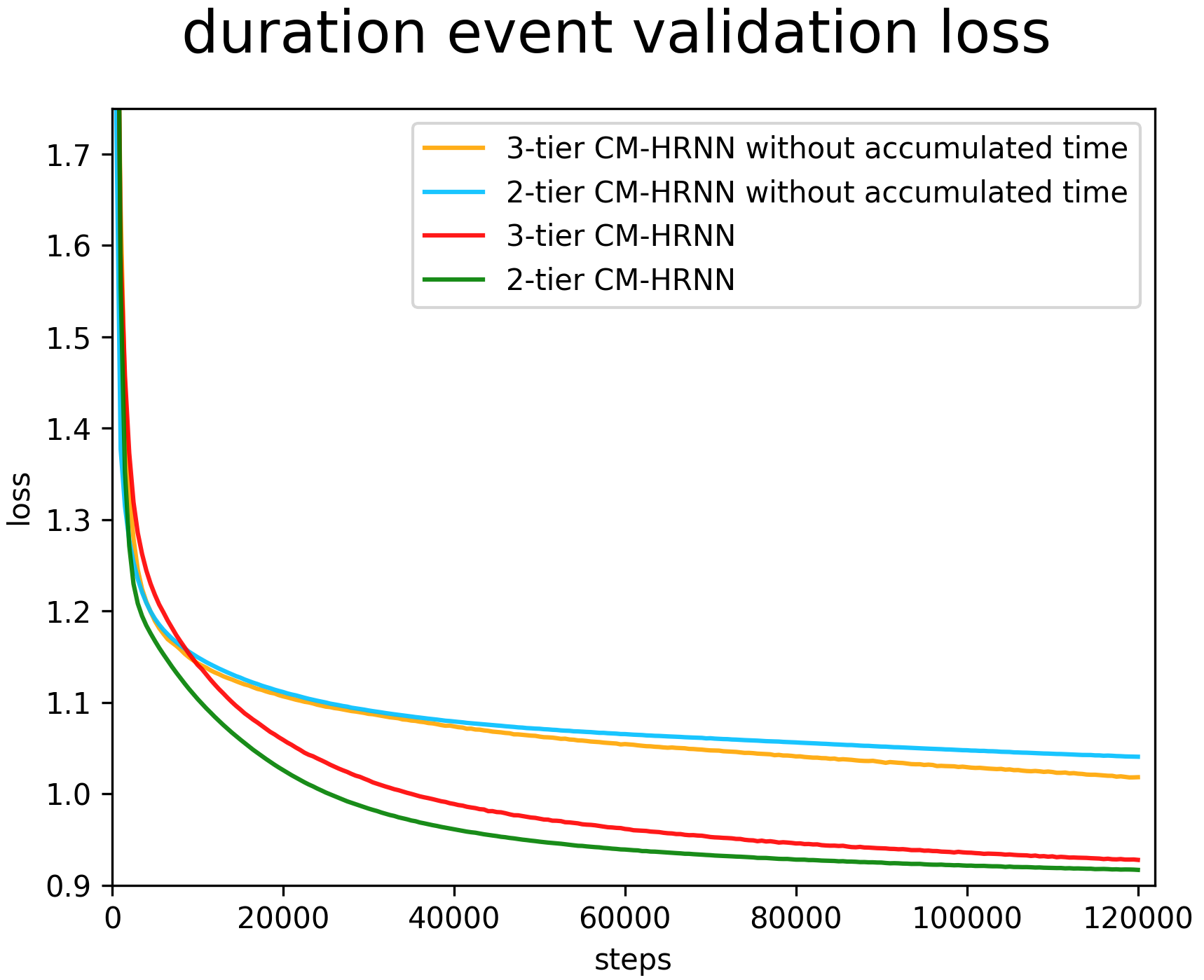}
		\caption{Duration event validation loss.}\label{fig:1b}
	\end{subfigure}\\
	\caption{Evaluation of the validation loss of different models, with and without added accumulated time information. Model configurations are shown in Table \ref{tab:sucessful_ratio}}\label{fig:bar_and_rhy_loss}
\end{figure}

To evaluate the influence of adding the accumulated time information in the bottom generation tier on the predicted duration of each note, as well as the correct bar placement, we trained 4 variants of our model (see Table \ref{tab:sucessful_ratio}): a model with 2-tiers and 3-tiers, each with and without the accumulated time information added as input. Intuitively, if the model can accurately predict when a new bar starts, and thus partly understands meter, we hope that it is better able to generate meaningful rhythm. The validation loss for models with and without accumulated time information is shown in Fig. \ref{fig:bar_and_rhy_loss}. It is obvious that by adding the accumulated time information, the models' validation loss, both when predicting the bar as well as the duration event, is lower than the models without the accumulated time information.

Additionally, we calculate the ratio of successfully predicted bar events among all predicted bars (successful bar ratio), $ \frac{no.\; of\; bars\; with\; 4\; beats}{total\; no.\;of\;bars}$ from the results generated by each model. We also compare the compression ratios of the 550 melodies generated by each model using the chord conditions of the test set as input. The results for each model configuration are shown in Table \ref{tab:sucessful_ratio}.  When comparing the successful bar ratios between models with and without added accumulated time information, we see that the former are almost always able to accurately predict when a new bar should start. In other words, they are able to partly understand musical meter. Also, models with accumulated time information achieve a much higher compression ratio compared to those without the accumulated time information, which indicates that the former can generate music with more repeated patterns.
% \begin{table}[h!]
% \centering
% \begin{tabular}{ |c|c|} 
%  \hline
% model no & successful bar ratio \\ \hline
% 1 & 88.4\% \\ \hline
% 2 & 91.2\%  \\ \hline
% 3 & 94.5\%\\ \hline
% 4 & 99.9\%  \\ \hline
% \end{tabular}
% \caption{sucessful bar ratio}
% \label{tab:sucessful_ratio}
% \end{table}

\begin{table}[h!]
\caption{Results of models with and without accumulated time information.}
\centering
\begin{tabular}{ lccccc } 
\toprule
Model&$FS^2$ &$FS^3$ &$acc_t$ & SBR &CPR \\  \midrule
2-tier CM-HRNN &16 &n.a. & no & 91.0\% & 1.61 \\
2-tier CM-HRNN &16 &n.a. & yes & \textbf{96.1\%} & \textbf{1.69}\\ 
3-tier CM-HRNN &2 &16 & no & 66.2\% & 1.62\\ 
3-tier CM-HRNN &2 &16 & yes & \textbf{100.0\%} & \textbf{1.68}\\ \bottomrule

\end{tabular}

  \smaller
  \smallskip
$acc_t$: accum. time information; SBR: successful bar ratio; CPR: compression ratio

\label{tab:sucessful_ratio}
\end{table}

\subsection{Residual connections and $FS^2$ in 3-tier CM-HRNN}
To validate the effectiveness of adding residual connections in the 3-tier CM-HRNN, we implemented six model variants, some with and without residual connections. To find the optimal $FS^2$ in a 3-tier model setting, we froze $FS^3$ to be 16 and experimented with different $FS^2$. The model configurations and results are shown in Table \ref{tab:diff_fs2}. From the results, given the same $FS^2$, it is clear that the residual connection can increase the overall compression ratio of the generated melodies. This is an indication that the generated music might contain more repeated themes and might have a larger overall structure \cite{chuan2018modeling}. Meanwhile, the successful bar ratio remains high for all 3-tier models regardless of $FS^2$. From these results, we choose the 3-tier model with $FS^2=2, FS^3=16$, and with residual connection as our best performing model.
\begin{table}[h!]
\caption{Comparing different $FS^2$ in a 3-tier model.}
\centering
\begin{tabular}{ ccccc} 
 \toprule
$FS^2$&$FS^3$&Residual connections&SBR&CPR\\ \midrule

2 &16  &yes & 100.0\% &\textbf{1.68}\\ 
2 &16  &no & 100.0\% &1.67\\ 
4 &16  &yes & 100.0\% &\textbf{1.64}\\ 
4 &16  &no & 100.0\% &1.63\\ 
8 &16  &yes & 99.2\% &\textbf{1.65}\\ 
8 &16  &no & 98.7\% &1.63\\ 
\bottomrule
\end{tabular}

  \smaller
% w\_res: with residual connection; 
\smallskip
SBR: successful bar ratio; CPR: compression ratio

\label{tab:diff_fs2}
\end{table}

\subsection{Comparison with AttentionRNN}
We compared our best performing model with a state-of-the-art model AttentionRNN~\cite{waite_2016}. This comparison was facilitated by the fact that their input representation is also multi-hot encoded, even though their original representation includes repetitive labels which indicate whether the note event was repeated 1 or 2 bars ago. Secondly, our problem domain is identical: modeling long-term coherence in music generation. Even though other research may share some similarities in terms of model architecture~\cite{roberts2018hierarchical,wu2019hierarchical}, they either work on a different problem domain or with different data representation, thus making comparison hard. 

We implemented two AttentionRNN models with the same LSTM setting: 2 layers (each with 256 nodes). One model has an attention lookback size of 16 and the other has an attention lookback size of 32. Other than the compression ratio and successful bar ratio, we also compare the tension of the generated music from all these models. We invite the reader to listen to some generated pieces online\footnote{\label{note1}\url{https://github.com/guozixunnicolas/CM-HRNN}}. 

%We show 3 generated melodies, one by 3-tier CM-HRNN, 2-tier CM-HRNN, and AttentionRNN with an attention lookback size of 32 respectively in Fig \ref{fig:comp}. The first two bars are the seed melody and all chords are from the guitar introduction of ``Sweet child o' mine''.

% \begin{figure}
%  \centering
% \includegraphics[scale=0.3]{figures/3t_2t_attn.png}
%  \caption{Generation results by CM-HRNN and AttentionRNN.}
%  \label{fig:comp}
%  \end{figure}

A comparison between the calculated measures for generated pieces by both our proposed CM-HRNN and AttentionRNN is shown in Table \ref{tab:comp_attnrnn}. From these results, we see that music generated by CM-HRNN has a much higher compression ratio compared to AttentionRNN. The tension values indicate that music generated by AttentionRNN sounds more tense and dissonant, with less tonality movement.

\begin{table}[h!]
\caption{Analytical measures for both CM-HRNN and AttentionRNN.}
\centering \smaller \setlength{\tabcolsep}{4pt}
\begin{tabular}{ lccccc } 
 \toprule
 &Meter&Structure&&Tension& \\ \midrule
Model&SBR&CPR& CD& TS& CM\\ \midrule
CM-HRNN(3t)      &100.0\% &\textbf{1.68} &\textbf{2.16$\pm$0.47} & 0.54$\pm$0.20& 0.69$\pm$0.21\\ 
CM-HRNN(2t)      &96.1\%  &\textbf{1.69} &2.17$\pm$0.47 &\textbf{0.53$\pm$0.20} &\textbf{0.70$\pm$0.21}\\ 
AttentionRNN(16) &88.2\%  &1.58 &2.22$\pm$0.47 &0.68$\pm$0.20 &0.52$\pm$0.20\\ 
AttentionRNN(32) &90.0\%  &1.58 &2.21$\pm$0.47 &0.68$\pm$0.20 &0.52$\pm$0.20\\ 
\bottomrule
\end{tabular}
\smaller
\smallskip
\\SBR: successful bar ratio; CPR: compression ratio; \\CD: cloud diameter; TS: tensile strain; CM: cloud momentum
\label{tab:comp_attnrnn}
\end{table}

\subsection{Listening test}
A total of 41 participants participated in the listening test. The resulting ratings in Table \ref{tab:listeningtst} show that our proposed CM-HRNN outperforms AttentionRNN in terms of overall enjoyment rating and long-term coherence, which again proves the effectiveness of our proposed model in capturing the long-term structure of music data.

\begin{table}[h!]\setlength{\tabcolsep}{4pt}
\caption{Listening test rating results on a scale from 1 to~5.}
\centering \smaller
\begin{tabular}{lccccc } 
 \toprule
&Overall && & Pitch& Duration\\
Model& Rating&Coherence& Consonance&naturalness & naturalness\\\midrule
CM-HRNN (3t) &\textbf{3.42} &\textbf{3.32} &\textbf{3.52} & \textbf{3.55}&3.39\\ 
CM-HRNN (2t) &3.35 &3.24 &3.49 & 3.52&\textbf{3.40}\\  
AttentionRNN (32) &2.80 &2.85 &2.95 & 2.98&2.79\\ 
\bottomrule
\end{tabular}
% \smaller

%cons:consonance; pn: pitch naturalness; dn: duration naturalness
\label{tab:listeningtst}
\end{table}

\section{Conclusion}
We propose a novel conditional hierarchical RNN network, CM-HRNN, to generate melodies conditioned with chords. In addition to using a novel, effective event-based representation that explicitly encodes bar information, CM-HRNN generates musically sound melodies that contain long-term structure. The TensorFlow implementation of the CM-HRNN implementation is available online\footnote[2]. 
In extensive experiments, both by using calculated features as well as a listening test, we show that pieces generated by CM-HRNN have greater tonal stability and more repeated patterns than those generated by AttenionRNN. 

%In future research, a custom loss function might be developed to ensure that the model is able to predict multiple correct notes (ex. C, E, G could be played under the context of a C major chord) instead of sampling a single correct note from the softmax probability distribution. Also, to further enhance the overall structure, instead of generating music note by note autoregressively, music events might also be generated in different hierachies.

\small
\bibliographystyle{IEEEtran}
\bibliography{References.bib}

\vspace{12pt}

\end{document}